\documentclass[Physsubmission, Phys]{SciPost}

\hypersetup{
    colorlinks,
    linkcolor={red!50!black},
    citecolor={blue!50!black},
    urlcolor={blue!80!black}
}

\usepackage[bitstream-charter]{mathdesign}
\DeclareSymbolFont{usualmathcal}{OMS}{cmsy}{m}{n}
\DeclareSymbolFontAlphabet{\mathcal}{usualmathcal}

\newcommand{\Msun}{M_\odot}
\newcommand{\Mstar}{M_\star}
\newcommand{\Rstar}{R_\star}

\newcommand{\mbeff}{m_i^{\rm eff}}

\newcommand{\fMB}{f_{\rm MB}}
\newcommand{\fFD}{f_{\rm FD}}

\newcommand{\sigmath}{\sigma_{th}}

\newcommand{\GeV}{{\rm \,GeV}}
\newcommand{\cm}{{\rm \,cm}}

\begin{document}
\begin{center}{\Large \textbf{
Improved Treatment of Dark Matter Capture in Compact Stars\\
}}\end{center}

\begin{center}
Sandra Robles 
\end{center}

\begin{center}
Theoretical Particle Physics and Cosmology Group, Department of Physics, King’s College London, Strand, London, WC2R 2LS, UK
\\
sandra.robles@kcl.ac.uk
\end{center}

\begin{center}
\today
\end{center}


\definecolor{palegray}{gray}{0.95}
\begin{center}
\colorbox{palegray}{
  \begin{tabular}{rr}
  \begin{minipage}{0.1\textwidth}
    \includegraphics[width=30mm]{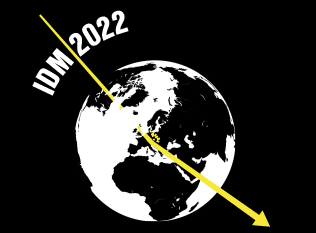}
  \end{minipage}
  &
  \begin{minipage}{0.85\textwidth}
    \begin{center}
    {\it 14th International Conference on Identification of Dark Matter}\\
    {\it Vienna, Austria, 18-22 July 2022} \\
    \doi{10.21468/SciPostPhysProc.?}\\
    \end{center}
  \end{minipage}
\end{tabular}
}
\end{center}

\section*{Abstract}
{\bf
Compact stellar objects 
are promising cosmic laboratories to test the nature of dark matter (DM). DM captured by the strong gravitational field of these stellar remnants transfers kinetic energy to the star during the collision. This 
can have various effects 
such as  anomalous heating of old compact stars. 
The proper calculation of 
the DM capture rate is key to derive bounds on DM interactions in any scenario involving DM accretion in a star. 
We improve former calculations, 
which rely on approximations,  
for both white dwarfs (WDs) and neutron stars (NSs). We account for the stellar structure, gravitational focusing,  relativistic kinematics,  
Pauli blocking,  
realistic form factors, and strong interactions~(NSs). 
Considering DM capture by scattering off either  ions or degenerate electrons in WDs, 
we show that old WDs in DM-rich environments 
could probe the elusive sub-GeV mass regime for both DM-nucleon and DM-electron scattering. In NSs, DM can be captured via collisions with strongly interacting baryons or relativistic leptons. 
We project the NS sensitivity to  DM-nucleon and DM-lepton scattering cross sections which greatly exceeds that of direct detection experiments, especially for low mass DM. 
}


\section{Introduction}
\label{sec:intro}
Direct detection (DD) experiments lead the quest to unveil the particle nature of dark matter (DM). 
In recent years, they have seen an impressive increase in sensitivity, especially to spin-independent (SI) interactions. However, their reach is limited by the achievable mass of the  target material and the recoil energy threshold. In addition, DD experiments are less sensitive to  
spin-dependent (SD) and DM-electron cross sections.
It is  then natural to look for alternative systems in which DM interactions lead to observable consequences. In this sense, DM capture  in the Sun  has long been used as an indirect detection technique. If DM couples to visible matter, it will scatter with the Sun  constituents. Provided that DM loses enough energy in the collision, it becomes gravitationally bound to the star. 
Accreted DM can be detected via  its annihilation to neutrinos that escape the Sun~\cite{Tanaka:2011uf, Choi:2015ara,Adrian-Martinez:2016gti,Aartsen:2016zhm,Bell:2021esh}.  

Because of their high density that will result in more efficient DM capture, compact stars were identified long ago as potential DM probes \cite{Goldman:1989nd,Kouvaris:2010jy}. It was recently pointed out that DM capture could transfer enough kinetic energy to  heat   old, isolated neutron stars (NSs) in the solar neighbourhood to infrared temperatures
\cite{Baryakhtar:2017dbj}. In light of this, 
in a series of papers 
we improved former  calculations of the DM capture rate, which rely on 
simplifying assumptions,  in both white dwarfs (WDs)~\cite{Bell:2021fye} and  NSs~\cite{Bell:2020jou,Bell:2020lmm,Bell:2020obw,Bell:2022,Anzuini:2021lnv}. We accounted for the stellar structure, gravitational focusing, a fully relativistic treatment of the scattering process, 
the star opacity, 
Pauli blocking  (for degenerate targets), 
 nuclear (WDs) and nucleon (NSs) form factors, and strong interactions (for baryonic targets in NSs). 
Using observations of old WDs in the globular cluster Messier~4 (M4)~\cite{Bedin:2009}, which we assumed to be formed in a DM subhalo, we derive bounds on DM-nucleon and DM-electron scattering cross sections. For NSs, we provide sensitivity projections to DM-nucleon and DM-lepton interactions, 
which surpass that of DD experiments  
especially for light DM. 
This paper is structured as follows. In section \ref{sec:compactstars}, we briefly summarise the internal structure of compact stars. In section~\ref{sec:capture}, we outline the capture rate calculation in both WDs and NSs. Our results are presented in section~\ref{sec:results} and concluding remarks in section~\ref{sec:conclusion}.

\section{Compact Stars}
\label{sec:compactstars}

The fate of a star is determined by its mass  when it enters the 
main sequence. 
Main sequence stars with masses below $\sim 8-10 \, \Msun$  end up their life cycles as WDs. 
More massive stars have a more spectacular end, a core-collapse supernova explosion that leaves behind a proto NS.

\subsection{White Dwarfs}
\label{sec:WDs}
WD progenitors are low and intermediate mass stars, therefore WDs are the most abundant stellar remnants. 
Moreover, WD physics is far more constrained than that of NSs. E.g., there is much less uncertainty 
in their equation of state (EoS), and their luminosity-age relation is  better understood. 
WDs are supported against gravitational collapse by electron degeneracy pressure. 
Most of them are  composed mainly of carbon and oxygen. 
To solve the WD  structure equations, we coupled the relativistic Feynman-Metropolis-Teller EoS~\cite{Rotondo:2009cr,Rotondo:2011zz}   with  
the  Tolman-Oppenheimer-Volkoff (TOV) equations~\cite{Tolman:1939jz,Oppenheimer:1939ne} (hydrostatic equilibrium in general relativity), and obtained  the WD mass $\Mstar$, radius $\Rstar$, as well as radial profiles of the ion $n_T(r)$ and electron $n_e(r)$ number densities, electron Fermi energy and escape velocity~$v_{esc}(r)$~\cite{Bell:2021fye}. 

\subsection{Neutron Stars}
\label{sec:NSs}
NSs are the most compact stars known in the Universe. Neutron degeneracy pressure supports them against collapse. Despite recent breakthroughs in NS physics, their exact composition  remains still unknown and the EoS of neutron-rich matter  an open problem in nuclear astrophysics. NSs are mainly composed  of degenerate neutrons, but inverse beta equilibrium allows the presence of protons and electrons. Muons appear in the NS core when the electron chemical potential reaches the muon mass. 
We model the NS interior and related microphysics by assuming a relativistic EoS that satisfies current observational constraints and enables the presence of hyperons in the NS inner core, the quark-meson coupling (QMC) model~\cite{Guichon:2018uew,Motta:2019tjc}, and solving the TOV equations. 
Radial profiles of the relevant quantities can be found in ref.~\cite{Anzuini:2021lnv}.

\section{Capture of Dark Matter in Compact Stars}
\label{sec:capture}

\subsection{Capture by scattering off ions}
\label{sec:capions}
First, we consider  DM scattering off the ionic targets in WDs. Since 
ions  are non-relativistic and the WD  gravitational potential is sufficiently weak so that Newtonian gravity holds, we use an approach similar to that of the Sun~\cite{Gould:1987ir,Gould:1987} to compute the capture rate~\cite{Bell:2021fye}
\begin{equation}
     C =
\frac{16\pi\mu_+^2\rho_\chi}{\mu m_\chi}\int_0^{R_\star} dr \,  n_T(r) \eta(r) r^2
     \int_0^\infty du_\chi \frac{\fMB(u_\chi)}{u_\chi}  \int_{w(r)\frac{|\mu_-|}{\mu_+}}^{v_{esc}(r)}dv v  \frac{d\sigma_{T\chi}}{d\cos\theta_{cm}} (w,q^2),\label{eq:capions}
\end{equation} 
where 
$\mu = m_\chi/m_T$, $\mu_\pm = (\mu\pm 1)/2$, 
$\rho_\chi$ is the DM density, $m_\chi$ the DM mass and $m_T$ the target mass, respectively; $\eta(r)$ is the optical factor that accounts for the star opacity, defined in refs.~\cite{Bell:2020jou,Bell:2021fye};  $w^2(r)=u_\chi^2+v_{esc}^2(r)$ and $v$ are the DM velocity before and after the collision, respectively; $q$ is the momentum transfer. 
We assumed a Maxwell Boltzmann distribution $\fMB(u_\chi)$ for the DM velocity far away from the star $u_\chi$. Note that  the differential DM-target cross section is written in the basis of non-relativistic operators \cite{DelNobile:2013sia} and includes the nuclear response function (form factors) as calculated in ref.~\cite{Catena:2015uha} (see ref.~\cite{Bell:2021fye} for further details).

\subsection{Capture by scattering off a free Fermi gas of degenerate leptons}
\label{sec:caplept}
Degenerate leptons in both, white dwarfs and neutron stars, are relativistic and subject to Pauli blocking. Therefore, Eq.~\ref{eq:capions}  cannot be applied to 
this case. 
We re-derived this expression using the TOV equations, the Schwarzschild metric and relativistic kinematics, and found~\cite{Bell:2020jou}
\begin{align}
C &= 4\pi \frac{\rho_\chi}{m_\chi} \int_0^\infty du_\chi \frac{\fMB(u_\chi)}{u_\chi}
\int_0^{\Rstar} dr \eta(r) r^2 \frac{\sqrt{1-B(r)}}{B(r)} \Omega^{-}(r)  , \label{eq:caprate}\\ 
\Omega^{-}(r) &= \frac{\zeta(r)}{32\pi^3}\int dt dE_i ds  \frac{ |\overline{M}(s,t,m_i)|^2}{s^2-[m_i^2-m_\chi^2]^2}\frac{ s E_i}{m_\chi}\sqrt{\frac{B(r)}{1-B(r)}}\frac{  \fFD(E_i,r)(1-\fFD(E_i^{'},r))}{\sqrt{(s-m_i^2-m_\chi^2)^2-4m_i^2m_\chi^2}}, 
\label{eq:intrate}
\end{align}
where 
$\fFD$ is the Fermi Dirac distribution, terms containing this function deal with  Pauli suppression of the target initial and final  states, $B(r)$ is the coefficient of the time part of the Schwarzschild metric and encodes general relativity corrections (very relevant for NSs),  $|\overline{M}|^2$, 
is the squared matrix element, $m_i$ is the mass of the target $i$, $s$ and $t$ are the  Mandelstam variables, $E_i$ and $E_i^{'}$ are the target  initial and final energies, respectively. The integration range for $s$, $t$ and $E_i$ can be found in refs.~\cite{Bell:2020jou,Bell:2021fye}. $\zeta(r)=n_i(r)/n_{free}(r)$ is a correction factor that accounts for the fact that  we are using realistic number density $n_i(r)$ and Fermi energy profiles while assuming a free Fermi gas. The expression for $n_{free}$  is given in ref.~\cite{Bell:2020jou}. 

\subsection{Capture by scattering off a Fermi sea of interacting baryons}
\label{sec:capbar}
At the extreme densities found in NSs, nucleons, and in general  baryons undergo strong interactions. 
Strong many body forces are described in terms of relativistic scalar and vector mean fields in the QMC EoS. Under the former field, baryons develop an effective mass $\mbeff$, 
which decreases with increasing density. Thus, $\mbeff$, where $i$ denotes the specific baryon, is lower than the rest mass  in vacuum $m_i$ towards the NS centre, and can be as low as $\sim0.5m_i$ for nucleons~\cite{Bell:2020obw,Anzuini:2021lnv}. This entails that the ideal Fermi gas  is not a good approximation to calculate the DM-baryon interaction rate Eq.~\ref{eq:intrate}. 
Properly incorporating the  effect of strong interactions in Eq.~\ref{eq:intrate} implies not only replacing $m_i$ with $\mbeff$, but also calculating the  Fermi energy of a single baryon 
as a function of its number density and $\mbeff$, and thereby 
 $\zeta(r)=1$~\cite{Bell:2020obw,Anzuini:2021lnv}. 
 
In addition, since DM is accelerated to quasi-relativistic speeds upon infall to a NS, the momentum transfer in the DM-baryon scattering process is sufficiently large  that baryon targets cannot be treated as point-like particles. We take this into account by incorporating the momentum dependence of the hadronic matrix elements. Thus, the squared    couplings of the baryon $i$ are $c_i(t) = c_i(0)/(1-t/Q_0^2)^4$, 
where  $Q_0\simeq1\GeV$ is a  scale 
that depends on the specific interaction and target, and $c_i(0)$ are  the squared coefficients at zero momentum transfer which depend on the hadronic matrix elements of the  specific interaction and baryon as in DD~\cite{Bell:2020obw,Bell:2022,Anzuini:2021lnv}.  
Note that the $t$-dependent baryon couplings are embedded in the squared matrix element $|\overline{M}(s,t,\mbeff)|^2$.

\section{Results}
\label{sec:results}

\begin{figure}[t]
\centering
\includegraphics[width=\textwidth]{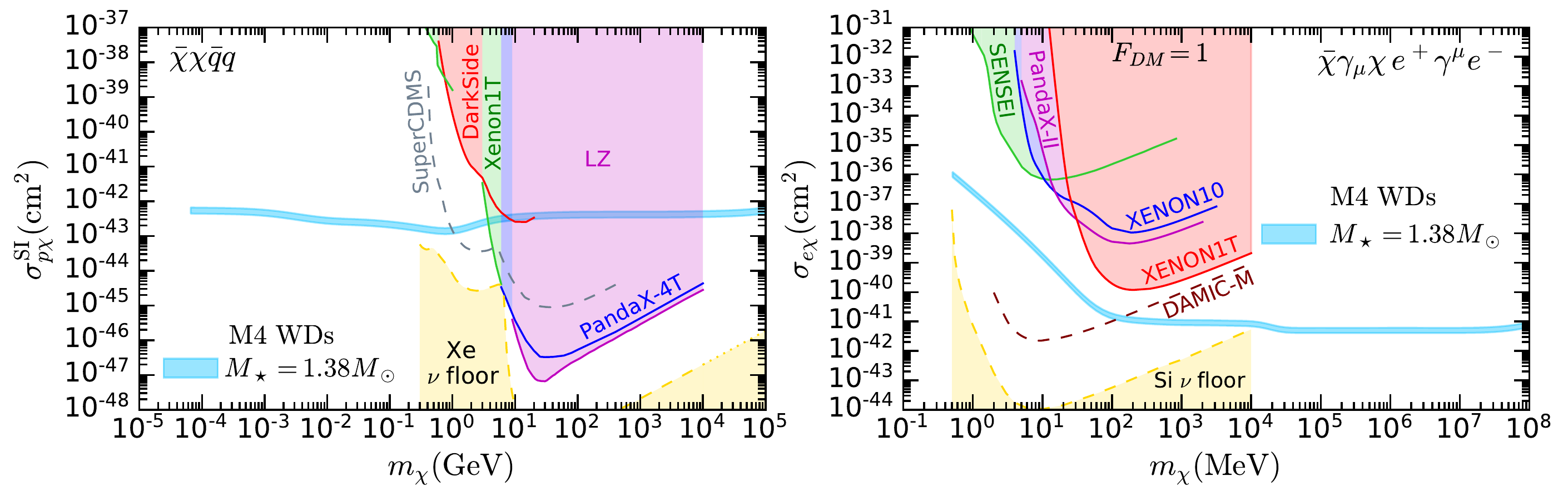}
\caption{Upper bounds (light blue band) on the DM-proton (left) and  DM-electron (right) scattering cross sections from WDs in the globular cluster M4, for the scalar and vector  operators, respectively;  assuming the existence of DM in M4. The band width depicts the uncertainty
in  $\rho_\chi$ in M4~\cite{McCullough:2010ai}. The leading  DD bounds, 
sensitivity projections from future experiments, and the neutrino floor~\cite{SENSEI:2020dpa,Essig:2017kqs,PandaX-II:2021nsg,Aprile:2019xxb,Agnese:2017jvy,Agnes:2018ves,Aprile:2019dbj,Aprile:2019jmx,Aprile:2020thb,PandaX-II:2016wea,PandaX-4T:2021bab,LUX-ZEPLIN:2022qhg,Essig:2015cda,Agnese:2016cpb,Ruppin:2014bra,Essig:2018tss} are also shown. }
    \label{fig:Wdbounds}
\end{figure}

We consider fermionic DM that scatters off either electron or ion targets in WDs, these interactions are described by the dimension-6 effective field theory (EFT) scattering operators~\cite{Bell:2021fye}. We compute the capture rate for carbon WDs using Eq.~\ref{eq:capions} for ions and Eqs.~\ref{eq:caprate} and \ref{eq:intrate} for electron targets, and 
the radial profiles obtained in section~\ref{sec:WDs}. 
Next, we derive limits on the cutoff scale of these operators by comparing the DM contribution to the WD luminosity due to capture and further annihilation with the observed luminosity 
of old 
WDs in the globular cluster M4~\cite{Bedin:2009}. 
The most constraining WD being the heaviest $\Mstar=1.38\Msun$  and faintest. 
Note that we have assumed  the existence of DM in 
M4,  which is yet to be proved, and 
$\rho_\chi\simeq531.5-798\GeV\cm^{-3}$ \cite{McCullough:2010ai}. In Fig.~\ref{fig:Wdbounds}, we recast these bounds in terms of the DM-proton (left panel) and DM-electron (right panel) cross sections (light blue band) for the scalar and vector operators, respectively. For DM-nucleon scattering, we find that
WDs  can probe the sub-GeV mass range, with its reach limited by evaporation~\cite{Bell:2021fye}. For DM-electron scattering, the WD bound  outperforms electron recoil experiments in the full mass range, with its low mass endpoint limited by DM annihilation to neutrinos that escape the WD~\cite{Bell:2021fye}.  

\begin{figure}[t]
\centering
\includegraphics[width=0.5\textwidth]{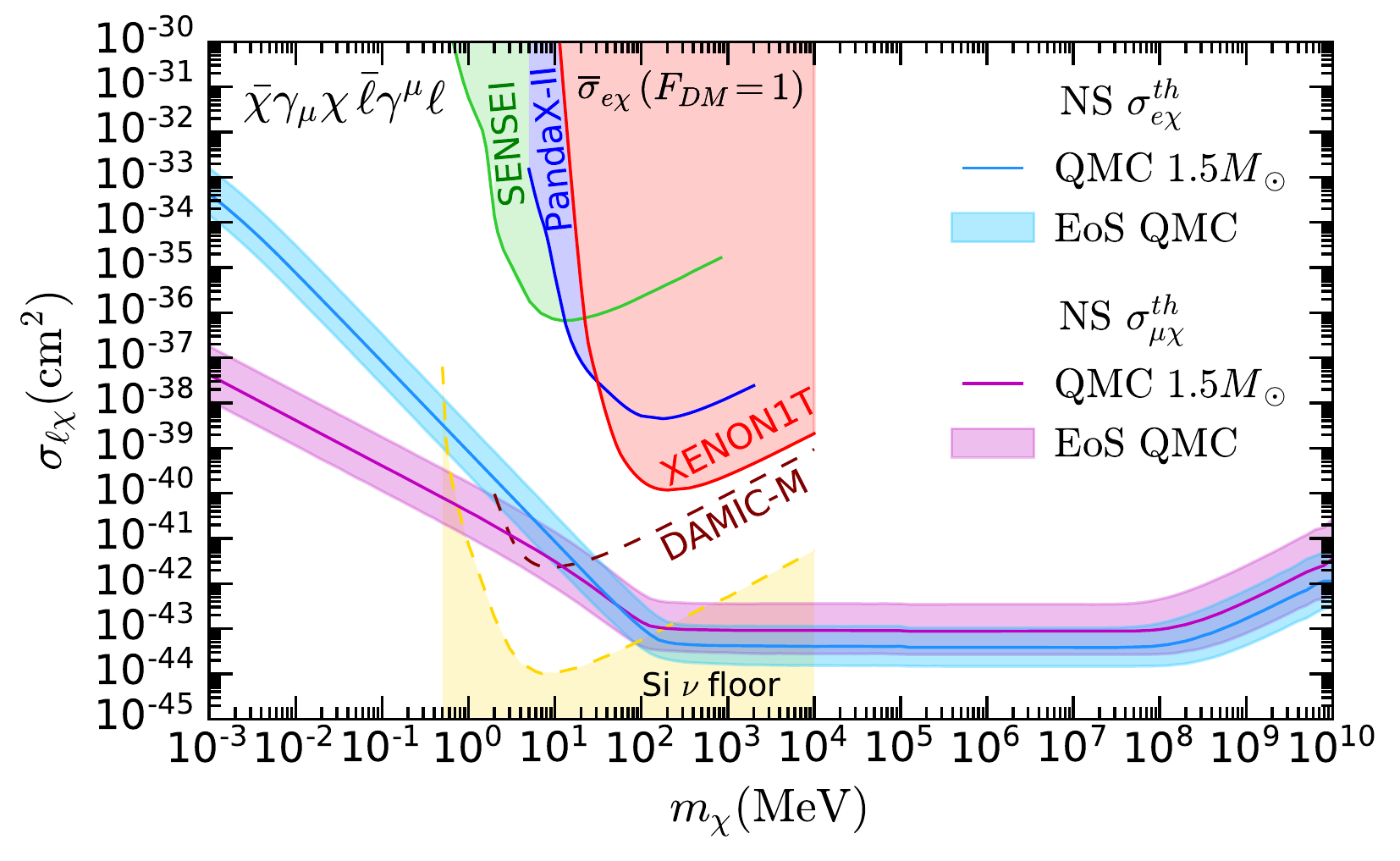} \\
\vspace*{-0.15cm}
\includegraphics[width=\textwidth]{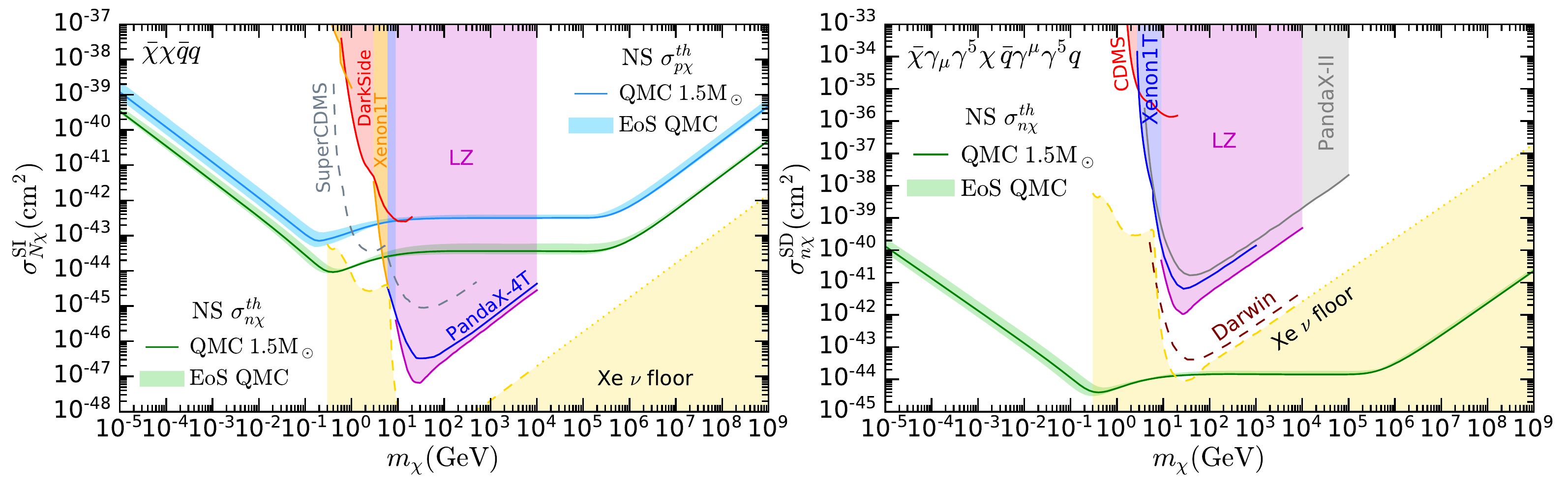}
\caption{Top: NS  sensitivity to DM-electron (light blue) and DM-muon  (magenta)  scattering  cross sections for the vector operator. 
Bottom: NS sensitivity to DM-neutron
(green) and DM-proton (light blue) interactions for
the  scalar (left panel) and axialvector (right panel) EFT operators. 
The solid  lines represent the  threshold cross section for a $1.5\, \Msun$ NS with a QMC EoS, and the shaded bands  the  variation in $\sigmath$ due to the EoS. 
We also show the leading  DD bounds, 
sensitivity projections from future experiments, as well as the neutrino floor~\cite{SENSEI:2020dpa,Essig:2017kqs,PandaX-II:2021nsg,Aprile:2019xxb,Agnese:2017jvy,Agnes:2018ves,Aprile:2019dbj,Aprile:2019jmx,Aprile:2020thb,PandaX-II:2016wea,PandaX-4T:2021bab,LUX-ZEPLIN:2022qhg,Essig:2015cda,Agnese:2016cpb,Ruppin:2014bra,Essig:2018tss}. 
}
\label{fig:sigmath}
\end{figure}

To project the NS sensitivity  to DM-nucleon and DM-lepton  scattering cross sections, we 
calculate the capture rate in the optically thin limit, $\eta(r)=1$, 
for the EFT operators, 
as outlined in sections~\ref{sec:caplept} and \ref{sec:capbar}, 
and the radial profiles from  section~\ref{sec:NSs} for NSs of mass in the $1-1.9\Msun$ range. 
To determine the   maximum cross section that can be probed with NSs,  the threshold cross section $\sigmath$, we equate $C
(m_\chi,\sigma_{i\chi})$ with the  expression for the geometric limit given in refs.~\cite{Bell:2018pkk,Bell:2019pyc}. 
Note that for $\sigma_{i\chi}>\sigmath$, the capture rate saturates the geometric limit.  
In Fig.~\ref{fig:sigmath}, we show $\sigmath$ for the vector operator and leptonic targets (top panel), as well as nucleon targets for the scalar (bottom left panel) and axialvector (bottom right panel) operators. The decrease in sensitivity below $m_\chi\sim0.2\GeV$ is due to Pauli blocking and that above $m_\chi\simeq4\times10^5\GeV$ (nucleons) and $m_\chi\sim[1,2]\times10^5\GeV$ (leptons) to the fact that multiple collisions are required to capture heavy DM. 
As we can see, the NS sensitivity greatly surpasses that of DD in the whole DM mass range considered for DM-neutron SD and DM-lepton interactions. 
The leading SI DD bounds are more stringent 
in the $\sim10-10^4\GeV$ mass range, below which the  NS sensitivity outperforms present and future DD experiments.

\section{Conclusion}
\label{sec:conclusion}

The extreme conditions found in compact stars made them promising 
dark matter (DM) probes.  
DM that accumulates and annihilates in the  interior of old isolated white dwarfs (WDs),  may transfer enough energy to these stars that can prevent them from cooling,  provided that they are located in DM-rich environments. Thus, the null detection of anomalously warm old WDs 
could  constrain DM interactions with ordinary matter. 
In neutron stars (NSs), on the other hand,  due to their stronger gravitational field that accelerates DM to quasi-relativistic speeds, only the energy transferred in the capture process would be enough to heat local NSs up to infrared temperatures for maximal capture efficiency. We have shown that the NS sensitivity 
excels that of direct detection experiments for DM-nucleon spin-dependent  and DM-lepton scattering
in the full DM mass range, and  for the  spin-independent scattering of sub-GeV  DM.  

\section*{Acknowledgements}
SR was supported by the UK STFC grant ST/T000759/1.
SR  thanks the  Institute for Nuclear Theory at the University of Washington for its hospitality and the Department of Energy for partial support during the completion of this work.

\bibliography{references.bib}

\nolinenumbers

\end{document}